\documentclass[a4paper]{article}
\usepackage{epsfig,amsmath,amsfonts,amssymb,setspace,multirow,textcomp}
\usepackage[T1]{fontenc}
\makeatletter
\renewcommand{\@oddhead}{\textit{Advances in Astronomy and Space Physics} \hfil}
\renewcommand{\@evenfoot}{\hfil \thepage \hfil}
\renewcommand{\@oddfoot}{\hfil \thepage \hfil}
\makeatother

\renewenvironment{thebibliography}[1]{\begin{oldthebibliography}{#1}\setlength{\parskip}{0ex}\setlength{\itemsep}{0ex}}{\end{oldthebibliography}}

\begin{document}
\fontsize{11}{11}\selectfont 
\title{MgH lines in the spectrum of Arcturus}
\author{\textsl{I.\,O.~Kushniruk$^{1}$, Ya.\,V.~Pavlenko$^{2}$, B.\,M.~Kaminskiy$^{2}$}}
\date{\vspace*{-6ex}}
\maketitle
\begin{center} {\small $^{1}$Taras Shevchenko National University of Kyiv, Glushkova ave., 4, 03127, Kyiv, Ukraine\\
$^{2}$Main Astronomical Observatory of the National Academy of Sciences of Ukraine, 27 Akademika Zabolotnoho St., 03680 Kyiv, Ukraine\\
{\tt nondanone@gmail.com}}
\end{center}

\begin{abstract}
The synthetic spectra of MgH lines was computed for the grid of the model atmospheres and compared with observed spectrum of Arcturus. The parameters of the atmosphere of Arcturus $\log{g}=1.5$ for $T_{eff}=4300 K$ were found by using the minimization procedure of differences between observed and computed spectra and compared with results of other studies.\\
{\bf Key words:} stars: atmospheres, stars: fundamental parameters, stars: individual (Arcturus), line: profiles.
\end{abstract}

\section*{\sc introduction}
\indent \indent The red giant Arcturus ($\alpha$ Boo, K2III) is one 
of the brightest star on the sky. It is interesting for studying, because
its mass is close to the mass of Sun, but Arcturus is at the later 
stage of evolution out of the Main Sequence. In some sense Arcturus represents 
the future of our Sun. Arcturus is known as an excellent reference star for 
spectroscopic studies of red giants, therefore availability precise 
atmospheric parameters for this star is very important. The atmospheric 
parameters of Arcturus have been estimated by several investigators using a variety of techniques, for example, the PASTEL database of stellar parameters by
Soubiran et al. (2010),  lists 28 entries for Arcturus. \\
\indent\indent One of the common problems in the investigations of the cool stars is the difficulties of the estimations of the surface gravity. Balmer lines in the spectra of such stars are sensitive to changes of $T_{eff}$ but not to $\log{g}$. In such case it's possible to use lines of some moleculas, which are sensitive to $\log{g}$ changes. MgH is suitable for this purpose, but we need to know an abundances of magnesium and hydrogen, because MgH lines are very sensitive to them. In such case we must search the self-consistent solution for $T_{eff}$, $\log{g}$ and abundance of Mg and H using the iteration method \cite{book}.\\
\indent At least three different MgH line lists are known at the present (lists by Yadin at al. \cite{Yadin2012},  Weck et al. \cite{weck2003}, Kurucz \cite{kurucz}). Some of them describe spectra well, others - not. In our work we used MgH line list by R. Kurucz \cite{kurucz}, which is very suitable for describing the spectrum. \\
\indent We investigated a few tasks here: Can we use MgH lines to refine $\log{g}$ and effective temperature of Arcturus? How do MgH lines in the spectrum of Arcturus depend on $\log{g}$? How do MgH lines in the spectrum of Arcturus depend on $T_{eff}$? 

\section*{\sc the calculations} 
\begin{figure}[!h]
\centering
\begin{minipage}[t]{.45\linewidth}
\centering
\epsfig{file = 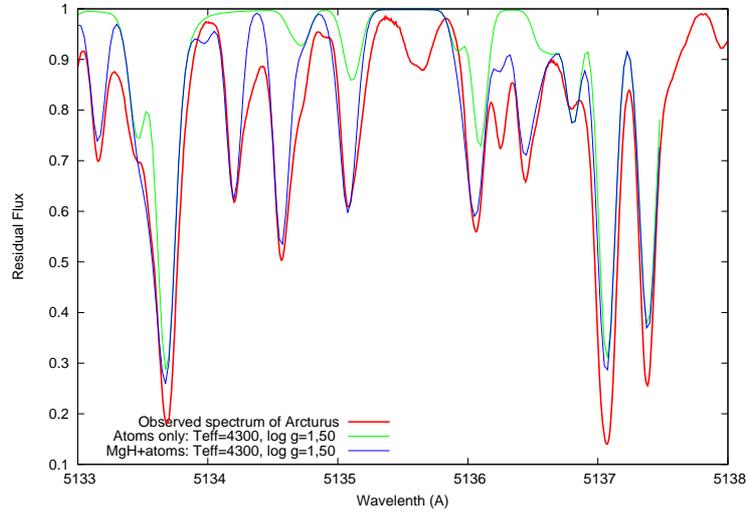,width = .85\linewidth, angle=270}
\caption{Red line shows observed spectra, green line is for convoluted synthetic spectra with an account of atomic lines, blue line shows convoluted synthetic spectra with an account of MgH and atomic lines in the range 5130-5140~\AA.}\label{fig1}
\end{minipage}
\end{figure}

\indent \indent  We computed 28 model atmospheres with different $\log{g}$ and $T_{eff}$: $\log{g}$ (0.75 - 2.25) with the step 0.25, $T_{eff}$ (4200 - 4500 K) with the step 100 by SAM12 program (Pavlenko, 2003) \cite{SAM12}. We computed synthetic spectrum of Arcturus by WITA6 \cite{WITA6}. Line list data were taken from database of atomic absorption spectra VALD (Kupka et al. 1999), observed spectrum from Visible and Near Infrared Atlas of the Arcturus Spectrum by Hinkle et al \cite{book}. For the first approach we used chemical composition found by PDK.  We accounted smoothing of the spectra of the instrumental profile, see Fig.~\ref{fig1}. We chose 23 "good" MgH features, which consist of more less clean (without atomic lines) MgH lines. Then we compared observed spectrum and computed synthetic spectrum of Arcturus. In order to do that, we computed deviation S between our MgH features in theoretical and observed spectra for all grid of the models, see formulas (\ref{form1}). 
\indent \indent 
\indent 

\begin{equation}\label{form1}
\sum_{i=1}^n \frac{(f_{{observed}_i}-f_{{computed}_i})^2}{l_2-l_1}=S_j,\\
\indent\indent\sum_{j=1}^N S_j=S,\\
\end{equation}
where $n$ is the number of points in the feature,\\
$N$ is number of features,\\
$S_i$ is a deviation for one of features,\\
$S$ is total deviation for the model atmosphere with some $\log g$ and $T_{eff}$,\\
$l_1$ and $l_2$ are boundary wavelengths of MgH feature.\\
\indent

\begin{figure}[!h]
\centering
\begin{minipage}[t]{.45\linewidth}
\centering
\epsfig{file = 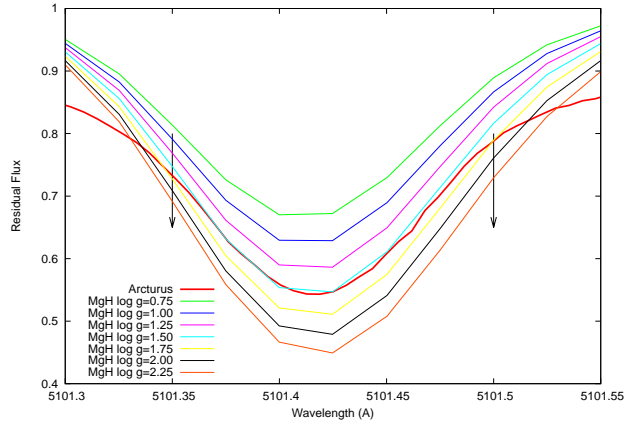,width = .7\linewidth, angle=270}
\caption{Dependence of residual flux on $\log{g}$ for $T_{eff}=4300 K$, (feature 1).}\label{fig2}
\end{minipage}
\end{figure}

The example of reseived dependence of residual flux on $\log{g}$ for one of 23 chosen lines is present on Fig.~\ref{fig2}  (here $T_{eff}$ is a constatnt and $T_{eff}=4300 K$).

\indent Arrows on the plots show borders of the feature $l_1$ and $l_2$.

\section*{\sc results and discussion}
\begin{figure}[!h]
\centering
\begin{minipage}[t]{.45\linewidth}
\centering
\epsfig{file = 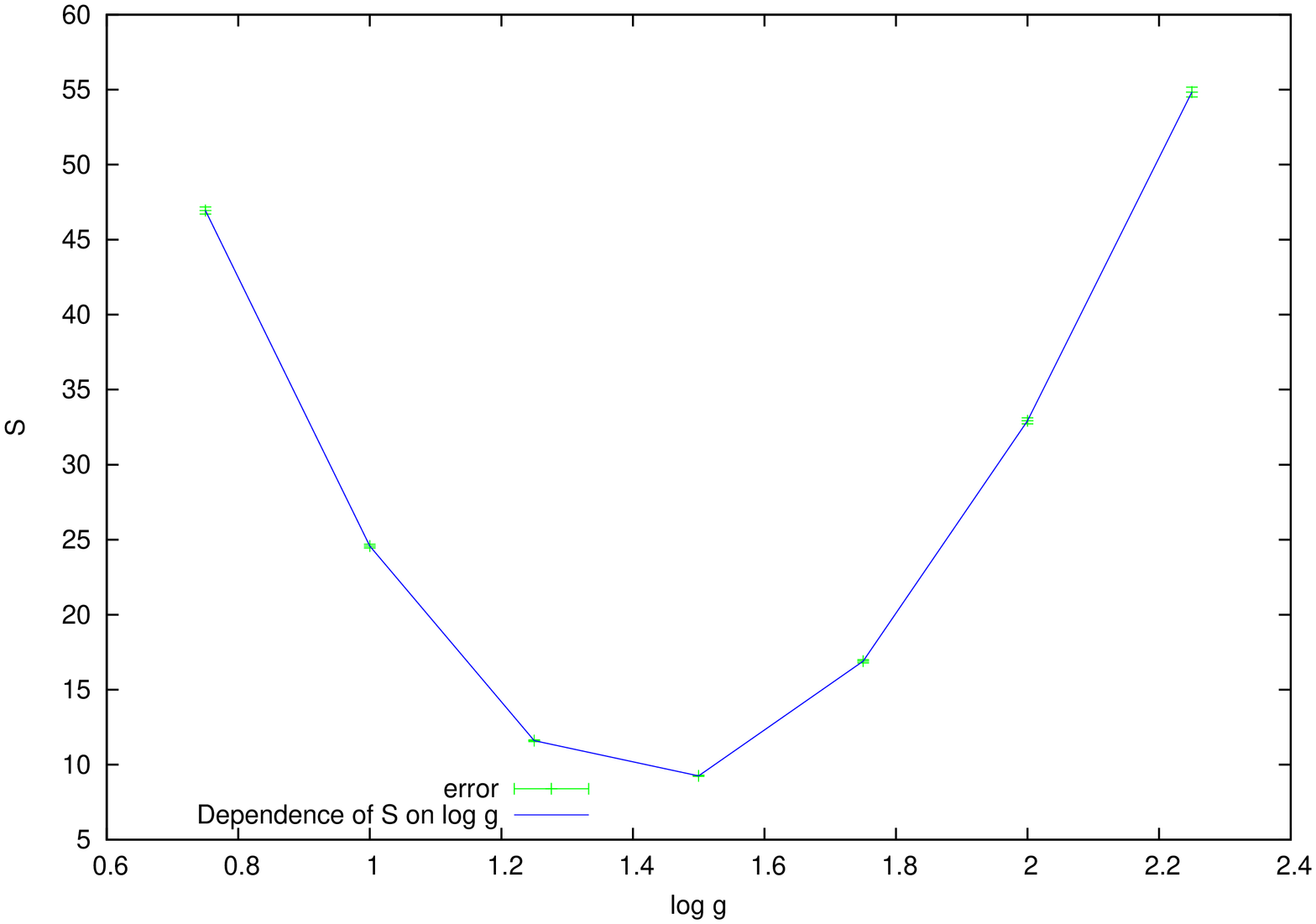,width = .7\linewidth, angle=270}
\caption{Dependence of $S$ on $\log{g}$ and $T_{eff}=4300 K$.}\label{fig3}
\end{minipage}
\hfill
\begin{minipage}[t]{.45\linewidth}
\centering
\epsfig{file = 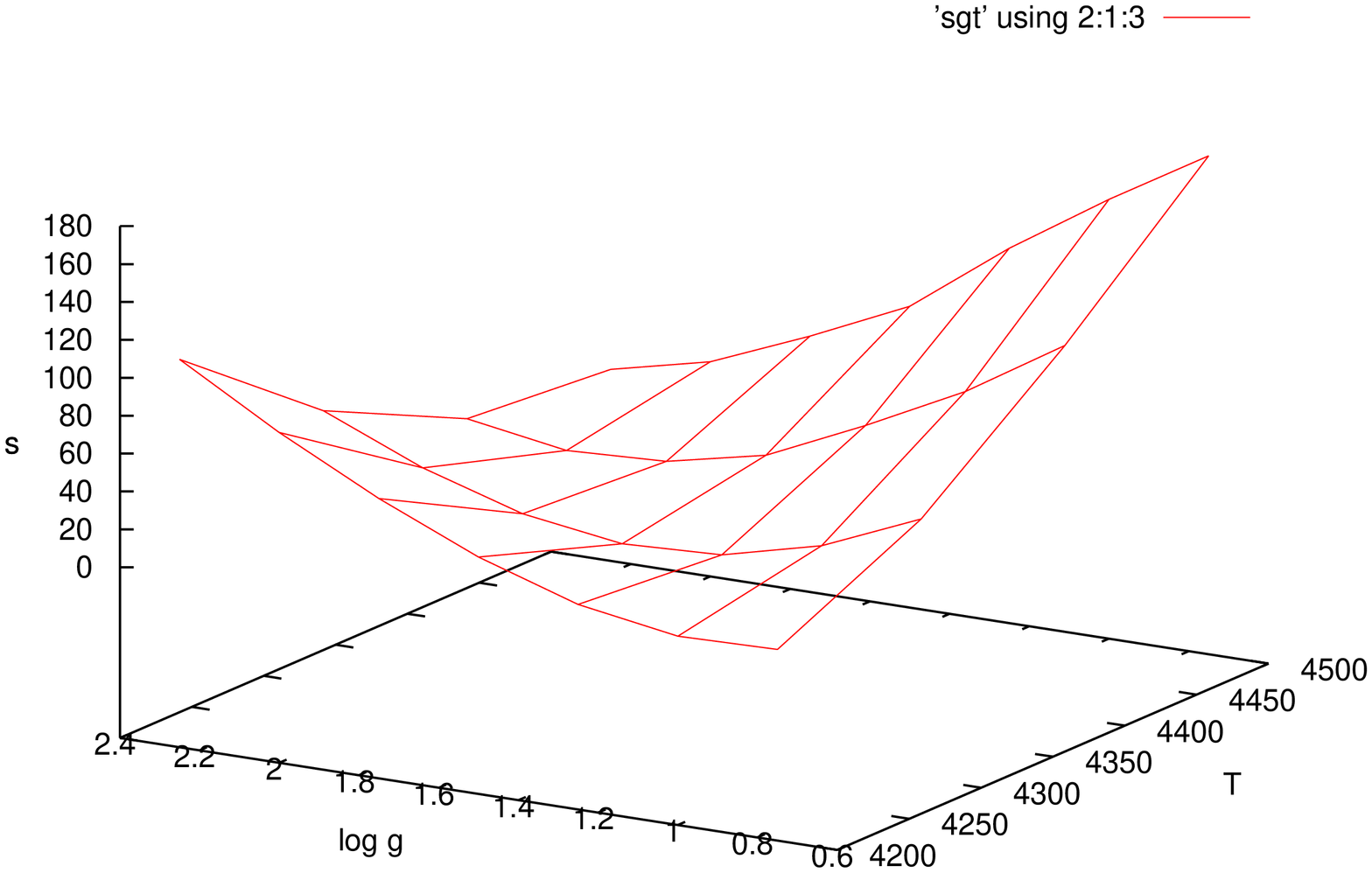,width = .7\linewidth, angle=270}
\caption{Dependence of $S$ on $\log{g}$ in $T_{eff}$.}\label{fig4}
\end{minipage}
\end{figure}

The results of the S value computations for all 23 lines in the case $T_{eff}=4300 K$ are shown in Table~\ref{tab1}, and Fig.~\ref{fig3}.
\begin{table}
 \centering
 \caption{Dependence of deviation on $\log{g}$ for $T_{eff}=4300 K$.}\label{tab1}
 \vspace*{1ex}
 \begin{tabular}{ccc}
  \hline
  $\log{g}$ & S & $\Delta{S}$ \\
  \hline
0.75 & 46.941200 & 0.2360\\
1.00 & 24.575607 & 0.1247\\ 
1.25 & 11.600469 & 5.7159E-002\\
1.50 & 9.2575245 & 3.9869E-002\\
1.75 & 16.896948 & 0.1063\\
2.00 & 32.928326 & 0.2056\\
2.25 & 54.842117 & 0.3246\\
 \hline 
 \end{tabular}
\end{table}

\begin{figure}[!h]
\centering
\begin{minipage}[t]{.75\linewidth}
\centering
\epsfig{file = 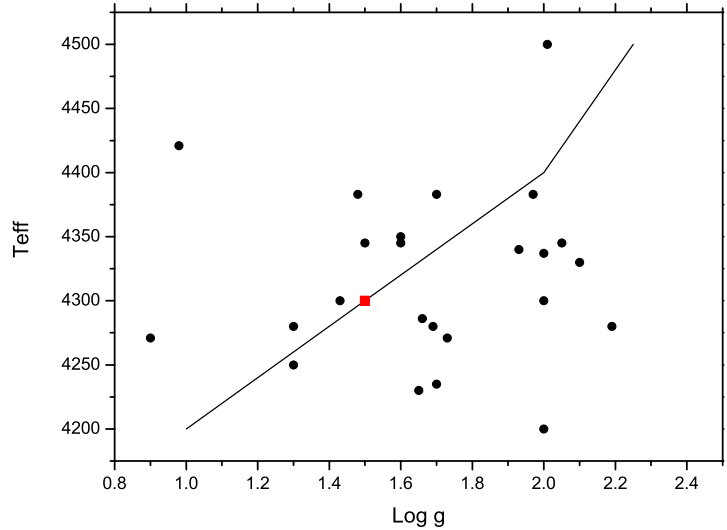,width = .7\linewidth} 
\caption{Projection of $S$ minimums (see Fig.~\ref{fig4}) on the plane $T_{eff}$-$\log{g}$, red dot is our value of fundamental parameters, black dots are values from other works mentioned in \cite{logg}.}\label{fig5}
\end{minipage}
\end{figure}

\indent 3D visualization of the dependences of $S$ on $T_{eff}$ 
and $\log{g}$ is on the plot (Fig.~\ref{fig4}),  we can find 
the min $S$ around $\log{g}=1.5$ for $T_{eff}=4300 K$. Here is 
exist clearly visible 
degeneration of $S$($T_{eff}$, $\log{g}$). In other words, $\log{g}$ can 
be determined if $T_{eff}$ 
is fixed by any other method. As Bell at al. (1985) noted,
a dominant source of errors of determination of log g using MgH lines is
the error in $T_{eff}$.\\

\indent The comparison of our result with results of 
other studies taken from the catalog Soubiran et al. (2010) is 
present on the plot (Fig.~\ref{fig5}). The line on the plot 
is the projection of minimums $S$ on the plane $T_{eff}$ vs. $\log{g}$, 
this line can be used as calibration curve for $\log{g}$ determination.
 Our values (red dot on (Fig.~\ref{fig5})) are close enough to the 
 middle of range, and at the same time are in excellent agreement with 
 results of PDK and some most recent works (see Soubiran et al. (2010) 
 for the references). There is another work \cite{loggo} were MgH lines
  was used for $\log{g}$ determination and give $\log{g}=1.8$ for effective 
  temperature of 4375 K. If we use our calibration curve for $T_{eff}=4375 K$
   we will obtain very close value of $\log{g}$, around 1.9. \\

\section*{\sc conclusions}

Our analysis was provided in the framework of classical approach. 
We did not account a few important processes witch might affect the 
intensities and profile shapes of spectral lines in stellar spectra, i.e.
NLTE, dependence of $V_t$ with depth, rotation, chromospheres 
(see Bell et al. 1985) for more details. Nevertheless, we obtained several 
interesting results of the common interest:  

\begin{itemize}

\item We selected 23 more less clean from the atomic 
lines MgH features in the wavelength range 5100-5200~\AA.
The information about them is available in the web: \\
(www.mao.kiev.ua/staff/yp/Results/MgHclean.ascii).

\item We show that absorption spectrum of MgH molecule is 
suitable for finding fundamental 
parameters of cool stars such as Arcturus.  

\item There is a degeneration of $S$ on $\log{g}$. We obtain 
the calibration curve which can be used for $\log{g}$ determination 
if $T_{eff}$ and abundance of Mg are fixed by any other independent method.

\item Our values for $T_{eff}$ and $\log{g}$ determined for Arcturus
 are in good 
agreement with other works.

\end{itemize}

\section*{\sc acknowledgement}
\indent \indent We thank Profs. Hinkle, Kurucz and VALD team for providing databases of astrophysical data used in our work,
Oleksiy Ivanyuk and Larisa Yakovina for the 
helpful comments, LOC and SOC of YSC 2013 for good organisation of the 
meeting and anonymous Referee for some positive remarks.


\end{document}